\documentclass[aps,pre,twocolumn,a4paper,10pt,notitlepage,footinbib,superscriptaddress,longbibliography]{revtex4-1}
\usepackage[english]{babel}
\usepackage{lmodern}
\usepackage[latin9]{inputenc}
\usepackage{endnotes}
\usepackage{amssymb,amsmath,amsfonts}
\usepackage{textcomp}
\usepackage{url}

\usepackage{graphicx}% Include figure files
\usepackage{dcolumn}% Align table columns on decimal point
\usepackage{bm}% bold math
\usepackage{xcolor}

\begin{document}

\title{Shear dynamics of polydisperse double emulsions}

\author{A. Tiribocchi}
\email{Corresponding author: adriano.tiribocchi@iit.it}
\affiliation{Center for Life Nano Science@La Sapienza, Istituto Italiano di Tecnologia, 00161 Roma, Italy}
\affiliation{Istituto per le Applicazioni del Calcolo CNR, via dei Taurini 19, 00185 Rome, Italy}
\author{A. Montessori}
\affiliation{Istituto per le Applicazioni del Calcolo CNR, via dei Taurini 19, 00185 Rome, Italy}
\author{F. Bonaccorso}
\affiliation{Center for Life Nano Science@La Sapienza, Istituto Italiano di Tecnologia, 00161 Roma, Italy}
\affiliation{Istituto per le Applicazioni del Calcolo CNR, via dei Taurini 19, 00185 Rome, Italy}
\affiliation{Department of Physics and INFN, University of Rome ``Tor Vergata'', Via della Ricerca Scientifica 00133 Rome, Italy.}
\author{M. Lauricella}
\affiliation{Istituto per le Applicazioni del Calcolo CNR, via dei Taurini 19, 00185 Rome, Italy}
\author{S. Succi}
\affiliation{Center for Life Nano Science@La Sapienza, Istituto Italiano di Tecnologia, 00161 Roma, Italy}
\affiliation{Istituto per le Applicazioni del Calcolo CNR, via dei Taurini 19, 00185 Rome, Italy}
\affiliation{Institute for Applied Computational Science, John A. Paulson School of Engineering and Applied Sciences, Harvard University, Cambridge, Massachusetts 02138, USA}

\date{\today}

\begin{abstract}
  We numerically study the dynamics of a polydisperse double emulsion under a symmetric shear flow. We show that both dispersity and shear rate crucially affect the behavior of the innermost drops and of the surrounding shell. While at low/moderate values of shear rates the inner drops rotate periodically around a common center of mass triggered by the fluid vortex formed within the emulsion generally regardless of their polydispersity, at higher values such dynamics occurs only at increasing polydispersity, since  monodisperse drops are found to align along the shear flow and become approximately motionless at late times. Our simulations also suggest that increasing polydispersity favours close-range contacts among cores and persistent collisions,  while hindering shape deformations of the external droplet. A quantitative evaluation of these effects is also provided.
\end{abstract}

\maketitle

\section{Introduction}

Double emulsions are an example of a highly structured fluid made of emulsion drops, of size up to $100\mu$m, containing smaller drops dispersed inside \cite{weitz,datta2014,guzowski2015,vladi2017,clegg2016,azar2019,tiribocchi_nat,nawar2020,santos2020,brower2020,werner2021}. A typical example is represented by a water core surrounded by an oil layer and immersed in water (often called water/oil/water emulsions), while higher complex arrangements may include collections of either mono or polydisperse distinct water cores placed within a larger oil drop \cite{kim2004comparative,utada2005monodisperse,abate2009,zarzar2015,ding2019,lee2016,li2020}.

Such emulsions are usually manufactured by means of microfluidic devices using, for instance, a single or a two-step emulsification process which ensure a regular design combined to a large production rate \cite{vladi2017}. Their typical compartmental template is a highly suitable feature for synthesizing structured soft materials useful in a wide number of technological applications, ranging from pharmaceutics for the transport and release of drugs \cite{laugel2000modulated,cortesi2002production,pays2002,kim2004comparative,timin2017,mao2019}, to food science for the production of dietary foods \cite{comunian2014,musciol2017,oppermann2017}, up to material science, for the realization of droplet-based porous materials \cite{chung2012,costantini2014}. They have also served as a unique platform to study interactions among vesicles and cells encapsulated within acqueous droplets \cite{kaminski2016,choi2016}.  

While many efforts have been dedicated to the experimental design and realization of such double emulsions, only recently a number of theoretical studies have addressed their dynamics in microfluidic channels \cite{tao2013,wang2020,tiribocchi_nat} and in the presence of external forcings \cite{ha1999,wang,smith,chen1,chen2,tiribocchi_pof,tiribocchi_prf}. Capturing the out of equilibrium behavior of these systems is crucial to manufacture materials combining a prolonged mechanical stability to a highly structural homogeneity, properties often required, for example, to design tissue scaffolds for medical purposes or pharmaceutical compunds for drug release.

The mechanical properties of a double emulsion can be controlled, for instance, by modulating the viscosity of the fluid layer to harden or jel the system \cite{utada2005monodisperse,omi2003}, or by additioning suitable surfactants adsorbed onto the interfaces of the cores to minimize their coalescence \cite{kim2011one}. A decisive role is also played by long-range hydrodynamic interactions, especially when a double emulsion is subject to an external flow, a backdrop particularly relevant in microfluidic and rheological experiments.  Such hydrodynamic-driven effects mediated by the fluid can destabilize the thin fluid film formed between opposite interfaces in high internal phase emulsion, thus yielding to considerable shape deformations \cite{sbragaglia1,sbragaglia3,costantini2014,raven2009}, and foster collisions among cores \cite{tiribocchi_nat,tiribocchi_prf}, thus favouring droplet breakup and merging, phenomena that ultimately alter the rate of polydispersity of the mixture. Indeed, although highly monodisperse double emulsions are generally required for the realization of soft materials or in biomedical applications \cite{song2012}, polydispersity may raise, for example, from the breakup of the jet of dispersed fluid at the nozzle of the injection channel or from breakup processes occurring downstream \cite{sauret2012,shum2012}. In this context it is thus essential to investigate how polydispersity may affect mechanical properties and morphology of double emulsions in the presence of an external driving.

In this work we theoretically study, by means of lattice Boltzmann simulations \cite{succi1,kruger,tiribocchi_nat}, the dynamic response of polydisperse double emulsions in the presence of a shear flow. In previous works, such as the ones described in Refs. \cite{chen1,chen2,smith}, numerical simulations have been dedicated to investigate shape deformations and breakups induced by a shear flow in a {\it single-core} emulsion, while only a few ones, like those in Refs. \cite{wang,tiribocchi_pof,tiribocchi_prf}, have been extended to {\it monodisperse} multi-core emulsions. We consider the simplest realization of a polydisperse double emulsion, essentially made of two drops of different size encapsulated within a larger drop. The rate of polydispersity is defined in terms of the index $h=R_1/R_2$, where $R_1$ and $R_2$ are the radii of the large and the small core, respectively (see Fig.\ref{fig1}). The physics of the emulsion is described by a multiphase field model \cite{marenduzzo1,marenduzzo2,tiribocchi_pof,tiribocchi_nat}, in which a set of balance equations governs the time evolution of hydrodynamic fields (density and velocity of the fluids) defined by a coarse-grained procedure over the microscopic details \cite{degroot}. 

In spite of its essential design we observe a variety of behaviors in which polydispersity plays a critical role. Indeed our simulations show  that, by varying shear rate and polydispersity index, the dynamic behavior at late times is generally characterized by the existence of nonequilibrium states displaying either, like in previous works \cite{tiribocchi_pof,tiribocchi_prf}, a persistent periodic motion of the cores triggered by the fluid vortex formed within the emulsion or, alternatively, new stationary configurations in which either one or both cores remain essentially motionless. In addition, increasing polydispersity is found to hinder shape deformations of the external interface, and to concurrently favour close-range contacts and collisions among internal cores. Such effects are assessed by measuring the Taylor parameter and by approximately computing the distance between the fluid interfaces of different cores. 

The paper is organized as follows. In the next section we shortly account for the computational model, while in section III we show the numerical results. We start by describing the dynamic response of emulsions with a different polydispersity  at low shear rates, and afterwards we move on to discuss the cases at higher shear rates. A quantitive evalutation of shape deformations of the external fluid droplet observed by varying the shear rate also provided. Some final remarks and perspectives of further research conclude the paper.

\section{Method}

\subsection{Equations of motion}

Following Refs. \cite{marenduzzo1,tiribocchi_pof,tiribocchi_nat}, we model a double emulsion by using a hydrodynamic phase field approach, in which a set of scalar fields  $\phi_i({\bf r},t)$, $i=1,....,N$ (where $N$ is the total number of droplets), accounting for the density of each droplet at position ${\bf r}$ and time $t$, is coupled to a vector field ${\bf v}({\bf r},t)$ describing the fluid velocity.

The equilibrium properties of this system are encoded in a free energy density \cite{degroot,landau} 
\begin{equation}\label{freeE}
f= \frac{a}{4}\sum_i^N\phi_i^2(\phi_i-\phi_0)^2+\frac{k}{2}\sum_i^N(\nabla\phi_i)^2+\epsilon\sum_{i,j,i<j}\phi_i\phi_j,
\end{equation}
where the first term allows for the existence of two coexisting phases, whose minima are $\phi_i=\phi_0$ (with $\phi_0\simeq 2$) inside the $i$th droplet and $0$ outside.
The second term accounts for the interfacial properties of the mixture and the last one mimics the repulsive effects due to a surfactant solution adsorbed onto the droplet interfaces.
The parameters $a$ and $k$ are positive constants defining the surface tension $\sigma=\sqrt{8ak/9}$ and the width of the interface $\xi=2\sqrt{2k/a}$ \cite{kruger,cates1}, while $\epsilon$ is a further positive constant gauging the strength of the repulsive interaction due to presence of a surfactant solution adsorbed onto the interfaces of the droplets.

The dynamics of the phase fields $\phi_i({\bf r},t)$ is governed by a set of advection-relaxation equations
\begin{equation}\label{CH_eqn}
\frac{\partial\phi_i}{\partial t}+{\bf v}\cdot\nabla\phi_i=M\nabla^2\mu_i
\end{equation}
where $M$ is the mobility and $\mu_i$ is the chemical potential of the $i$th drop, given by
\begin{equation}
\mu_i=\frac{\partial f}{\partial\phi_i}-\frac{\partial_{\alpha}f}{\partial(\partial_{\alpha}\phi_i)}.
\end{equation}

The evolution of the density $\rho({\bf r},t)$ and fluid velocity ${\bf v}({\bf r},t)$ are controlled by the continuity and Navier-Stokes equations which, in the incompressible limit, are
\begin{equation}\label{CNT_eqn}
\nabla\cdot{\bf v}=0,
\end{equation}
\begin{equation}\label{NAV_eqn}
  \rho\left(\frac{\partial}{\partial t}+{\bf v}\cdot\nabla\right){\bf v}=-\nabla p +\eta\nabla^2{\bf v}-\sum_i\phi_i\nabla\mu_i.
\end{equation}
Here $p$ is the isotropic pressure and $\eta$ is the dynamic viscosity.

\subsection{Numerical details}

Eqs.~(\ref{CH_eqn}), (\ref{CNT_eqn}) and (\ref{NAV_eqn}) are solved by using a hybrid lattice Boltzmann (LB) approach, in which the advection-relaxation equations are integrated via a finite difference scheme while the continuity and the Navier-Stokes equations through a standard LB algorithm \cite{succi1,kruger,montessori,sukop}. In particular, the former (\ref{CH_eqn}) is solved by using an explicit Euler algorithm and the latter (\ref{CNT_eqn},\ref{NAV_eqn}) by means of a predictor-corrector scheme. This hybrid method has been extensively adopted to simulate binary fluids \cite{yeomans_pre,tiribocchi,tiribocchi2}, liquid crystals \cite{tiribocchi3}, pressure driven flows \cite{marenduzzo1,marenduzzo2,tiribocchi_nat} and active matter \cite{tiribocchi4} and, with respect to other models \cite{chen1,wang},  allows for the inclusion of a high number of immiscible droplets potentially containing orientational fluids confined within the shell \cite{nieves2007}.

Simulations are performed on two dimensional rectangular lattices of size $L_y=500$ (horizontal) and $L_z=250$ (vertical). Two parallel flat walls are placed at distance $L_z$, where neutral wetting holds for $\phi_i$ and no-slip conditions for ${\bf v}$. The former means that ${\bf n}\cdot\nabla\mu_i|_{z=0,L_z}=0$ (no flux through the boundaries) and ${\bf n}\cdot\nabla(\nabla^2\phi_i)|_{z=0,L_z}=0$ (interfaces perpendicular at the boundaries), where ${\bf n}$ is an inward normal unit vector at the walls, while the latter that $v_z(z=0,z=L_z)=0$.

\begin{figure*}
\includegraphics[width=1.\linewidth]{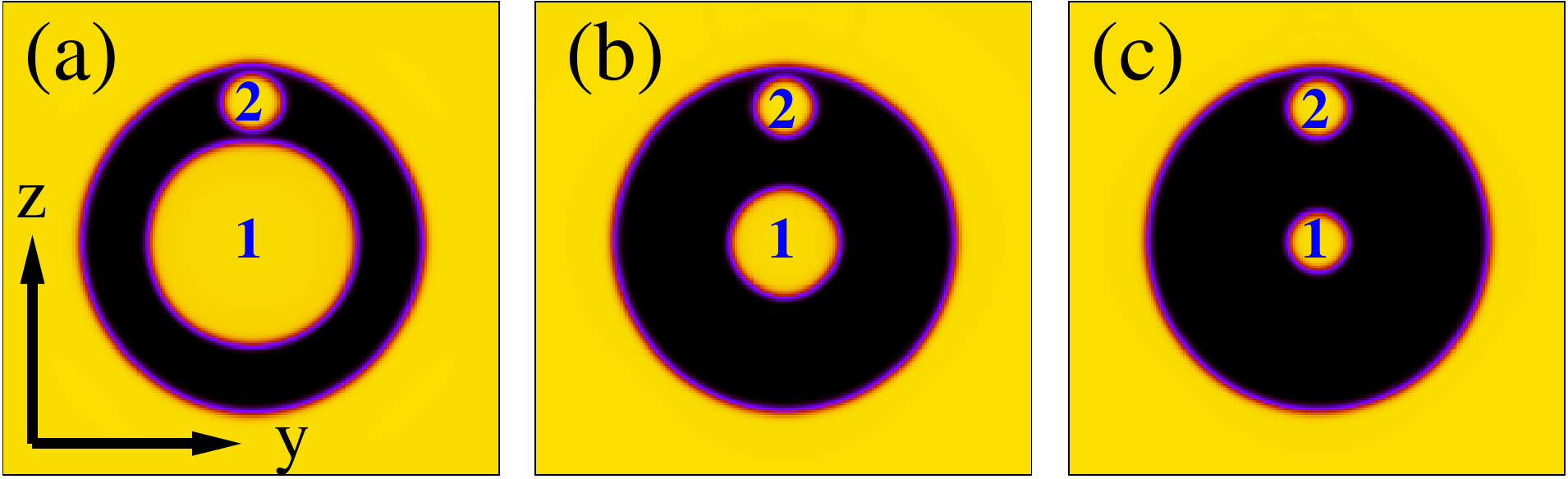}
\caption{Equilibrium configurations of three double emulsions with different size of the inner drops, (a) $R_1=40$, $R_2=10$ ($h=4$) (b) $R_1=20$, $R_2=10$ ($h=2$) and (c) $R_1=R_2=10$ ($h=1$). The radius of the external droplet has been kept fixed to $R_{3}=68$. Only a portion of the lattice (of full size $L_y\times L_z=500\times 250$) is shown. Colors correspond to the values of the order parameter $\phi$, ranging from $0$ (black) to $\simeq 2$ (yellow). This applies to all figures.}
\label{fig1} 
\end{figure*}

In Fig.\ref{fig1} we show three examples of double emulsions with different polydispersity, defined through the index $h=R_1/R_2$, where $R_1$ and $R_2$ are the radii of cores $1$ and $2$ respectively. In Fig.\ref{fig1}a the equilibrium radius of droplet $1$ is $R_1=40$ lattice sites while the one of droplet $2$ is $R_2=10$. This sets the polydispersity ratio $h=R_2/R_1=4$ and the area fraction $A_f=\frac{\pi\sum_iR_i^2}{\pi R_{3}^2}\simeq 0.37$, where $R_{3}=68$ is the radius of the external droplet (the black region in Fig.\ref{fig1}). In Fig.\ref{fig1}b $R_1=20$ and $R_2=10$ ($h=2$, $A_f\simeq 0.11$), and in Fig.\ref{fig1}c $R_1=R_2=10$ ($h=1$, $A_f\simeq 0.04$). In these configurations one hase $N=3$ phase fields, since three separate drops are considered. The fields $\phi_1$ and $\phi_2$ (associated to drops $1$ and $2$) are positive (equal to $\simeq 2$) within each drop and zero everywhere else, while the field $\phi_3$ (associated to the external droplet) is positive outside the larger droplet and zero elsewhere.

Droplets are initially relaxed to achieve a near equilibrium state and afterwards a symmetric shear is applied, by moving the top wall along the positive $y$-axis and the bottom wall along the opposite direction,  with velocity $v_w$ and $-v_w$ respectively. In our simulations $v_w$ ranges between $0.005$ (low shear) to $0.03$ (moderate/high shear), which means that the shear rate $\dot{\gamma}=2v_w/L_z$ varies between $\simeq 4\times 10^{-5}$ to $\simeq 2.5\times 10^{-4}$.

The evolution is expressed in terms of a dimensionless time $t^*=\dot{\gamma}(t-t_{eq})$ \cite{chen2,tiribocchi}, where $t_{eq}$ is the relaxation time after which the shear is switched on, approximately equal to $10^4$ time steps. The thermodynamic parameters have been chosen as follows: $a=0.07$, $k=0.1$, $M=0.1$, $\eta=1.67$, $0.01<A_f< 0.5$, $\Delta x=1$ (mesh step), $\Delta t=1$ (time step) and $\epsilon=0.05$. The latter, in particular, gauges the repulsive effect due to the presence of a surfactant, and is set to value sufficient to prevent droplet merging. These parameters define $\sigma\simeq 0.08$, $\xi\simeq 3.5$ and $D=Ma=0.001$ (where $D$ is the diffusion constant), values ensuring good numerical stability and close mapping the real units (see below).

Finally, the dynamic behavior is described by means of capillary and Reynolds numbers, defined as $Ca=\frac{v_{max}\eta}{\sigma}$ and $Re=\frac{\rho v_{max}L}{\eta}$, where $L$ is a characteristic length (such as the radius of the external droplet) and $v_{max}$ is the maximum speed measured. The former ranges between $0.1$ and $1$ and the latter between $1$ and $5$, thus breakup events are absent and turbulent-like effects are negligible. As in previous studies \cite{marenduzzo1,tiribocchi_pof,tiribocchi_nat}, fixing the length scale, time scale and force scale to $L=1\mu m$, $T=10\mu s$ and $F=100nN$, our system could be mapped onto a microfluidic channel of length $\sim 0.5-1$mm, in which drops of radius ranging from $\sim 10\mu$m to $\sim 100\mu$m, surface tension $\sigma \sim 1$mN/m and diffusion constant $D\simeq 10^{-10}m^2/s$ are surrounded by a Newtonian fluid of viscosity $\eta\sim 0.1-1$ Pa$\cdot$s \cite{mezzenga}. A typical speed of the cores is about $\sim 1$ $mm/s$ under a shear rate of $\sim 1/s$. With these numbers gravity effects can be neglected since, assuming $\Delta \rho/\rho_w=(\rho_w-\rho_o)/\rho_w\sim 0.1$ (where the water density $\rho_w=10^3Kg/m^3$ and a typical oil density $\rho_o\sim 9\times 10^2Kg/m^3$), one has a Bond number $Bo=\Delta\rho g R^2/\sigma\sim 10^{-2}\div 10^{-3}$, where $g$ is the gravity acceleration.

\section{Results}

Here we discuss our numerical results about the dynamic response of a polydisperse double emulsion under a symmetric shear flow. We initially describe the physics observed at low shear rates (which means $\dot{\gamma}\leq 6\times 10^{-5}$) and afterwards we move on to evalutate the effects produced by higher values of $\dot{\gamma}$.

\subsection{Low shear rates}

\begin{figure*}
\includegraphics[width=1.\linewidth]{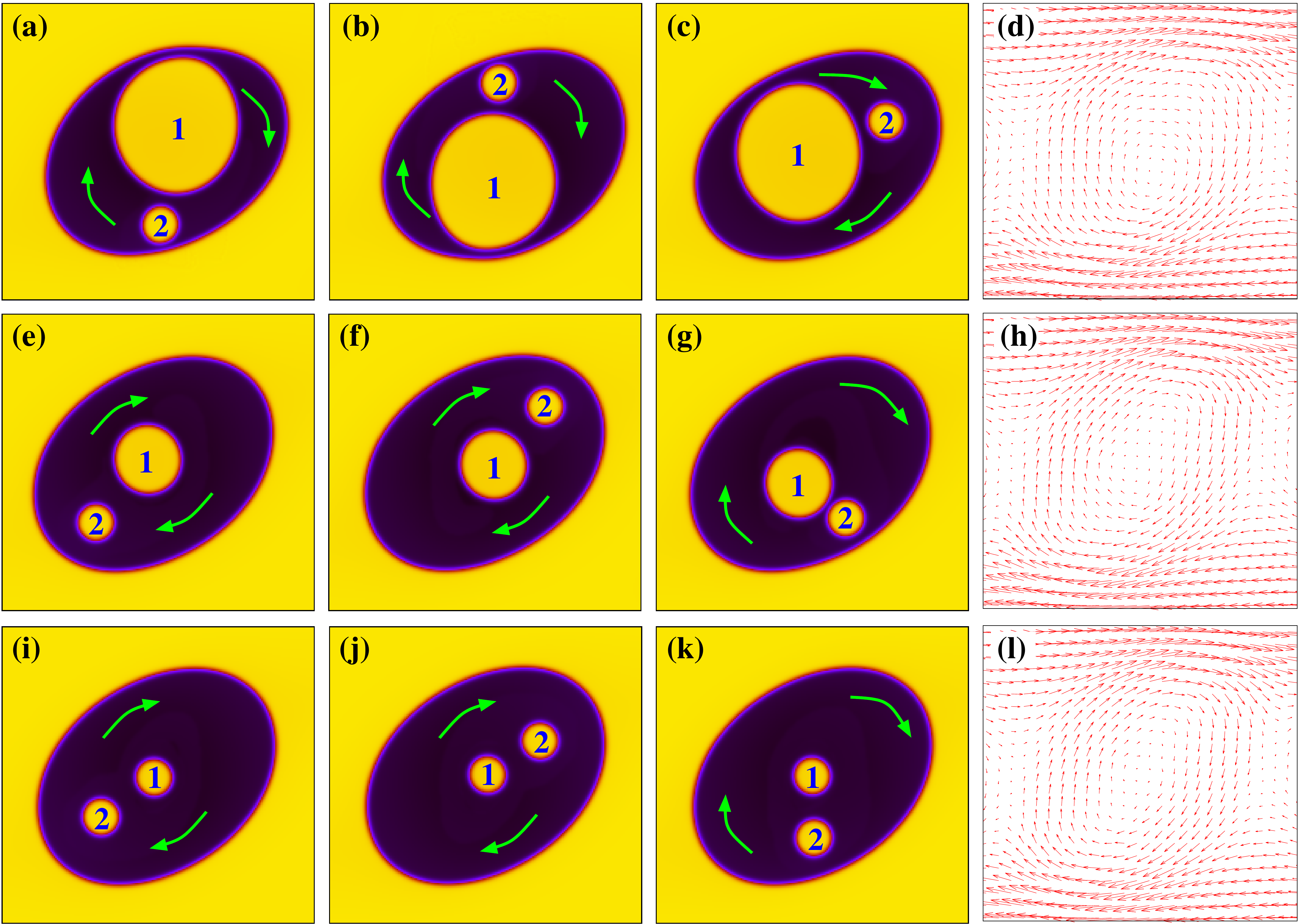}
\caption{Late-time dynamics of three different examples of polydisperse double emulsions under shear flow, with $\dot{\gamma}\simeq 6\times 10^{-5}$. In (a)-(c) we show three instantaneous configurations of the cores with $h=4$, in (e)-(g) with $h=2$ and in (i)-(k) with $h=1$. Here $Re\sim 0.6$ and $Ca\sim 0.15$. While if $h=4$ (a-c) both cores rotate periodically following elliptical paths, decreasing $h$ (e-g) gradually stabilizes the droplet in the center of the emulsion. If $h=1$ (i-k) the core in the middle becomes essentially motionless. Green arrows denote the trajectories follow by the droplets. Snapshots are taken at $t^*=24.6$ (a)-(e)-(i), $t^*=32.4$ (b)-(f)-(j) and $t^*=36.6$ (c)-(g)-(k). In (d), (h) and (l) we show the typical steady state fluid recirculation taken at $t^*=36.6$. Multimedia views: (a), (e) and (i).}
\label{fig2} 
\end{figure*}

\begin{figure*}
\includegraphics[width=1.\linewidth]{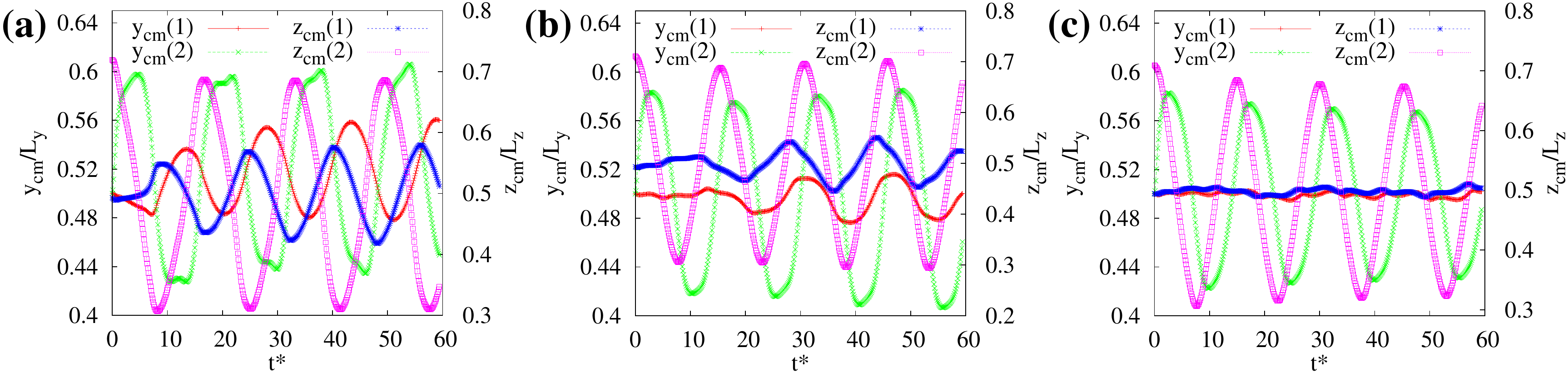}
\caption{Time evolution of the $y$ and $z$ components of the center of mass of the two cores for $h=4$ (a), $h=2$ (b) and $h=1$ (c), with $\dot{\gamma}\sim 6\times 10^{-5}$. The motion of the core located in the middle of the emulsion (red/plusses and blue/asterisks plots) varies from a periodic circular one for high values of $h$, to one in which it remains substantially at rest in that position for low values of $h$. The motion of the core placed near the external interface (green/crosses and purple/squares plots), on the contrary, follows a periodic circular trajectory essentially regardless of the polydispersity.}
\label{fig3} 
\end{figure*}

Starting from the equilibrated configurations reported in Fig.\ref{fig1}, we apply a symmetric shear flow by moving top and bottom walls along opposite directions. In Fig.\ref{fig2}a-d (Multimedia view) we show a typical late-time dynamics observed for polydispersity index $h=4$. This is in agreements with results shown in previous works \cite{tiribocchi_pof,tiribocchi_prf}, in which monodisperse droplets under shear flow have been found to exhibit a periodic planetary-like motion driven by a fluid vortex formed within the external droplet. Likewise, here the two innermost cores display a periodic clockwise motion (see Fig.\ref{fig3}a where the time evolution of the centers of mass is plotted) travelling along almost elliptical trajectories, without undergoing appreciable shape deformations due to the relatively mild magnitude of the velocity field.

Diminishing the polydispersity $h$ and the area fraction $A_f$ alters this behavior. In Fig.\ref{fig2}e-g (Multimedia view) we show the late-time dynamics of the polydisperse emulsion with $h=2$. Once again the fluid vortex (Fig.\ref{fig2}h) triggers and sustains the periodic motion of the smaller drop along a circular path, but weakly affects the position of the other core placed in the middle of the emulsion. Indeed its center of mass exhibits fluctuations markedly smaller than the ones shown by the large core (see Fig.\ref{fig3}b). This is essentially because the reduction of droplet size entails a decrease of the drag caused by the fluid in that region. Such an effect is even more emphasized with $h=1$ [Fig.\ref{fig2}i-l, (Multimedia view)], a case in which the drop located at the center is essentially at rest (Fig.\ref{fig3}c) while the other one ceaselessly moves around in an Earth-Sun fashion. 

It is worth mentioning that such dynamics importantly depends on the initial position of the internal cores. This is particularly true at low values of polydispersity (and low values of area fraction $A_f$), since, in these cases, the motionless core in the middle of the emulsion would easily acquire motion if placed off-center. We anticipate that this picture does not hold at higher shear rates, a scenario discussed in the next section.

\subsection{High shear rates}

We now consider the dynamic response of the polydisperse multiple emulsion shown in Fig.\ref{fig1}, under a symmetric shear flow with $\dot{\gamma}\simeq 2.5\times 10^{-4}$.

\begin{figure*}
\includegraphics[width=1.\linewidth]{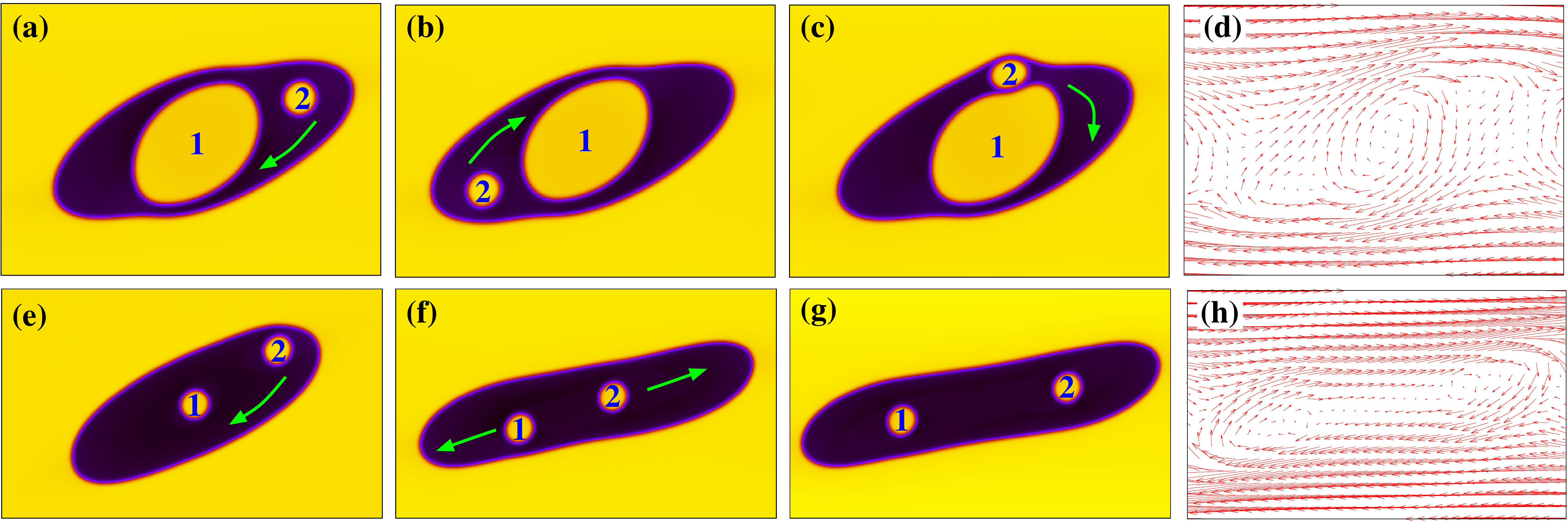}
\caption{Late-time dynamics of two polydisperse double emulsions under shear flow, with $\dot{\gamma}\simeq 2.5\times 10^{-4}$. In (a)-(c) we show three instantaneous configurations of the cores with $h=4$ and in (e)-(g) with $h=1$. Here $Re\sim 2.5$ and $Ca\sim 0.6$. In contrast with the dynamics observed at low shear rates, in (a-c) the small core moves around following an elliptical orbit, while the large one is only weakly shifted from the center of the emulsion. In (e-g) the system relaxes towards a state in which both cores stabilize on opposite sides, after a transient regime in which they weakly come into contact in the middle of the emulsion. Green arrows denote the trajectories followed by the droplets. Snapshots are taken at $t^*=38.4$ (a), $t^*=55.2$ (b), $t^*=69.6$ (c), $t^*=2.4$ (e), $t^*=26.4$ (f), $t^*=93.6$ (g). In (d) and (h) we show the typical steady state fluid recirculations taken at $t^*=36.6$. The mildly stretched structure observed in (d) is replaced by a highly elongated one in (h) containing two smaller vortices located within the cores. Multimedia views: (a) and (e).}
\label{fig4} 
\end{figure*}

\begin{figure*}
\includegraphics[width=1.\linewidth]{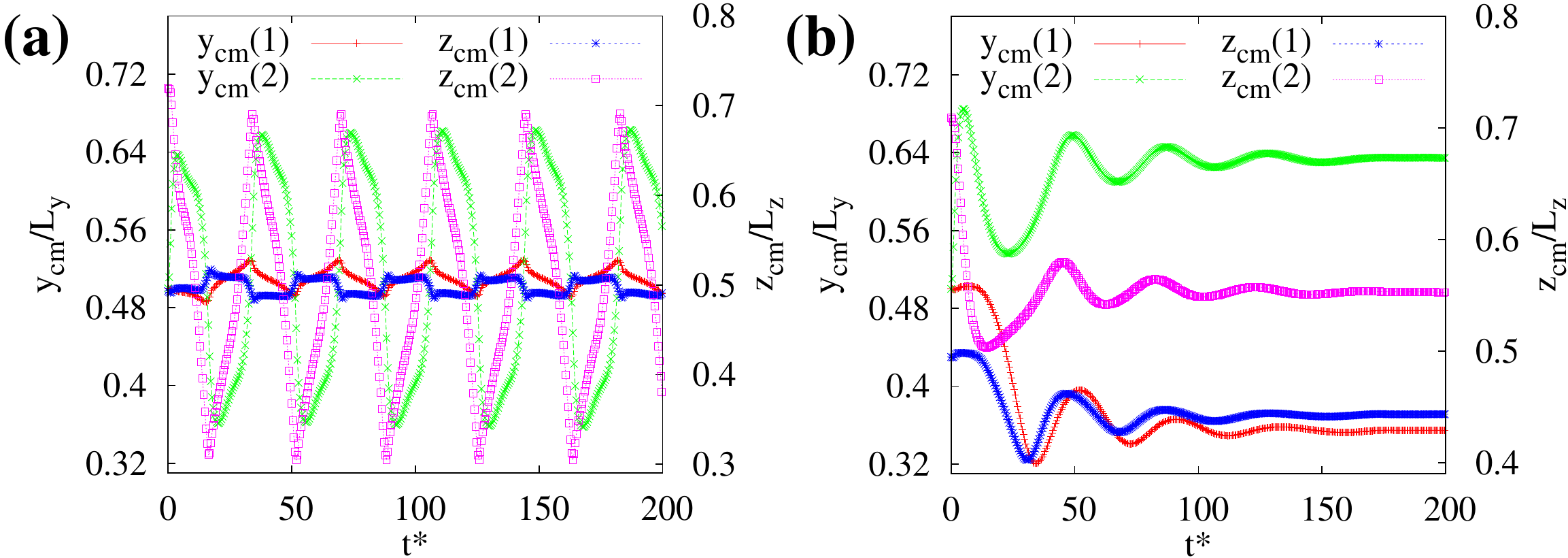}
\caption{Time evolution of the $y$ and $z$ components of the center of mass of the two cores for $h=4$ (a), $h=1$ (b), with $\dot{\gamma}\sim 2.5\times 10^{-4}$. If $h=4$ the smallest core (green/crosse and purple/squares plots) persistently rotates around the largest one (red/plusses and blu/asterisks plots), whose position weakly fluctuates in a periodic manner. If $h=1$ the two cores move towards opposite extremities of the emulsion and remain essentially motionelss at late times.}
\label{fig5} 
\end{figure*}

In Fig.\ref{fig4}a-d (Multimedia view) we show the typical nonequilibrium steady states and velocity field observed at $h=4$. Unlike the case at low shear rate, here the large core remains essentially in the middle of the emulsion, displaying periodic oscillations (see Fig.\ref{fig5}a) triggered by momentum transfer and repeating collisions of the small core. This one travels along clockwise elliptical trajectories and is capable of crossing through narrow fluid interstices (whose size is generally smaller than the one of the rotating core) formed between the interface of the large core and that of the external droplet. Such an effect is determined by the sheared structure of an intense velocity field, made of a large fluid recirculation located in the middle of the emulsion surrounded by a stretched fluid vortex, whose magnitude augments nearby the fluid interstices. This dynamics produces local interface bumps significantly affecting the shape of both the external droplet and the large internal one. In the next section we provide a quantitative evaluation of such deformations and we show that, at fixed shear rates, these ones can be reduced by increasing the polydispersity.

At lower values of $h$ [Fig.\ref{fig4}e-h, (Multimedia view)] the cores follow an alternative dynamics, in which they initially move away from the center of the emulsion and afterwards stabilize and arrange at approximately opposite sides within a highly elongated droplet. The shape of the latter results from the structure of the velocity field which, unlike the previous case, consists of an eccentric vortex containing two smaller fluid recirculations located within each core. Note that, at a given shear rate, a different polydispersity index considerably alters the steady state pattern of the velocity field. This is overall not surprising, since changing the index $h$ modifies the term ${\bf v}\cdot\nabla\phi$ appearing in Eq.\ref{CH_eqn} and governing the coupling between fluid velocity and phase field. However, as previously mentioned, this result also indicates that the shape deformations of outer drop observed under flow are significantly affected  by the polydispersity of the cores. This is precisely the point discussed in the next section.

\subsection{Shape deformations}

We dedicate this section to describe the shape deformations of the external droplet observed by varying $\dot{\gamma}$ and $h$.
As long as, under shear flow, the Reynolds number is below or approximately around $1$ and the capillary number remains lower than $1$ as well but higher than the critical value below which coalescence can occur, droplet breakups are very unlikely events and an emulsion generally attains, at the steady state, the shape of an ellipse. This result is rather general and holds for both single and double emulsions under a symmetric shear flow \cite{tiribocchi_pof,chen1}. A suitable number providing a quantitative estimate of such shape changes is the Taylor parameter defined as $D=\frac{a-b}{a+b}$, where $a$ and $b$ are the major and minor semiaxis (by definition we assume $a\ge b$). Its value ranges between $0$ (perfectly circular shape) to $1$ (needle-like shape).

\begin{figure*}
\includegraphics[width=1.\linewidth]{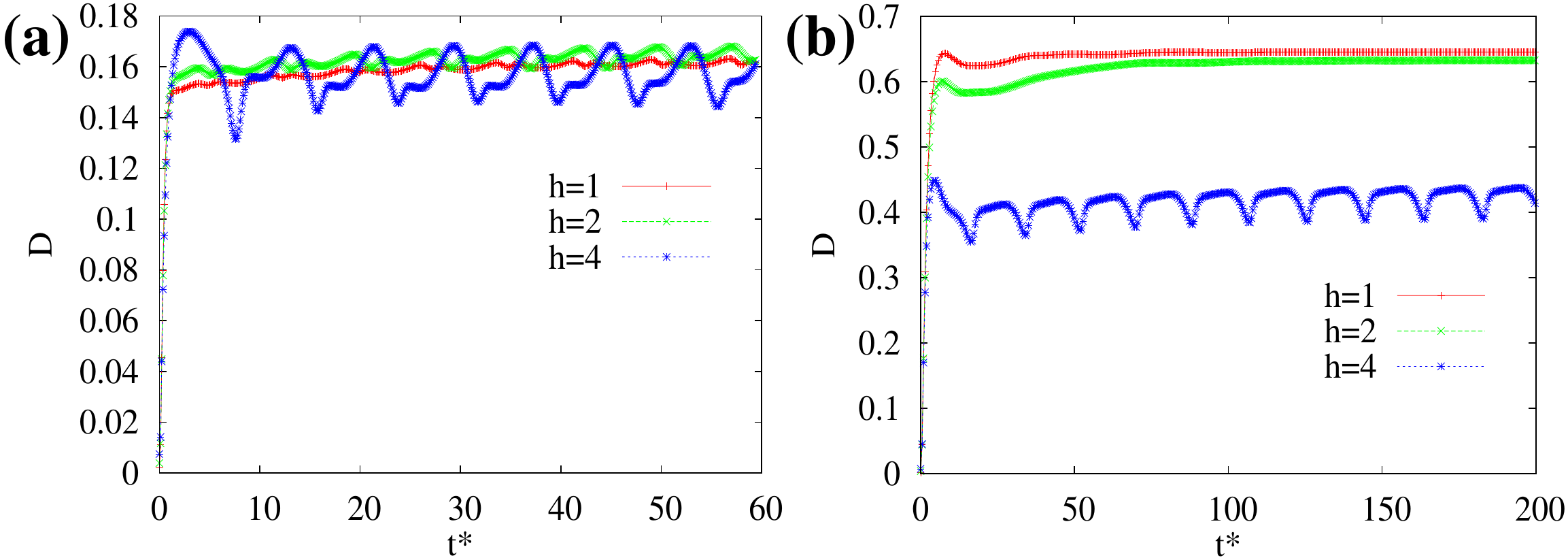}
\caption{Time evolution of the deformation parameter $D$ of the external droplet for $\dot{\gamma}\simeq 6\times 10^{-5}$ (a) and $2.5\times 10^{-4}$ (b). While in (a) $D$ attains approximately the same value regardless of the values of $h$, in (b) $D$ decreases for increasing $h$. This suggests that, at high shear rates, increasing the polydispersity of the cores diminishes the deformation of the external droplet. Time oscillations observed for $h=4$ are due to the recurring  periodic collsions of the two cores against the external interface.}
\label{fig6} 
\end{figure*}

In Fig.\ref{fig6} we show the time evolution of $D$ for $\dot{\gamma}\simeq 6.5\times 10^{-5}$ (Fig.\ref{fig6}a) and $\dot{\gamma}\simeq 2.5\times 10^{-4}$ (Fig.\ref{fig6}b), representative of low and high shear rates regimes. At low $\dot{\gamma}$, the value of $D$ looks independent on the polydispersity, although larger fluctuations emerge for increasing $h$. These are due to the persistent collisions of the rotating cores against the external interface, yielding to a periodic stretch-and-shorten dynamics of the drop. On the contrary, at high $\dot{\gamma}$ the deformation parameter significantly diminishes for increasing values of $h$, essentially because the large internal drop preserves a less stretched, approximately circular, shape which, in turn, prevents intense deformation of the external drop. Note also that $D$ fluctuates only for high values of $h$, while at low $h$ it keeps a constant value, since collisions of the cores against the outer interface are essentially negligible.

\subsection{Interactions and close-contact dynamics among cores}

The physics discussed so far suggests that interactions and contacts among cores can be favoured in systems with a higher polydispersity index. Controlling their reciprocal distance and position is certainly relevant in biomicrofluidic experiments, to detect, for example, the pathogenicity of bacteria, causing human diseases, in cells (the inner drops in this model) hosted within an acqueous droplet \cite{kaminski2016}. Such interaction can be estimated by computing the distance $|\Delta {\bf r}_{cm}|=|{\bf r}_{cm,1}-{\bf r}_{cm,2}|$ between the centers of mass of each core with respect to a distance $d=R_1+R_2+l$, where $R_1$ and $R_2$ are the radii of the cores and $l$ is the length of the thin film formed between opposite interfaces in close contact. If $|\Delta {\bf r}_{cm}|\leq d$ the cores are sufficiently close to temporarily sustain a thin fluid film which locally affects the shape of the interfaces,  whereas if $|\Delta {\bf r}_{cm}|> d$ drops are considered too far away and reciprocal interaction can be assumed negligible. Since the width of the interface is approximately $\xi \simeq 4\div 5$ lattice sites, we set $l=2\xi\simeq 10$, thus giving $d=30$ for $h=1$ ($R_1=R_2=10$), $d=40$ for $h=2$ ($R_1=20$, $R_2=10$) and $d=60$ for $h=4$ ($R_1=40$, $R_2=10$).

\begin{figure*}
\includegraphics[width=1.\linewidth]{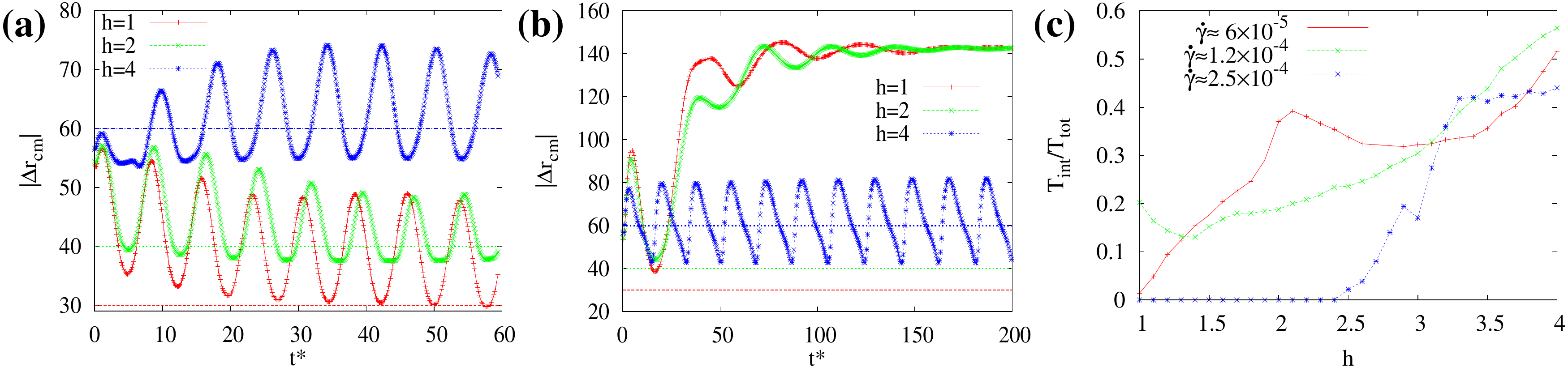}
\caption{(a)-(b) Time evolution of the distance $|\Delta {\bf r}_{cm}|=|{\bf r}_{cm,1}-{\bf r}_{cm,2}|$ between the centers of mass of both cores for $\dot{\gamma} \simeq 6\times 10^{-5}$ (a) and $\dot{\gamma}\simeq 2.5\times 10^{-4}$ (b), at $h=1$ (red/plusses), $h=2$ (green/crosses) and $h=4$ (blue/asterisks).  Horizontal lines indicate the distance $d=R_1+R_2+l$ below which opposite interfaces sustain a temporary fluid film. Here $l=2\xi$, where $\xi\simeq 4-5$ lattice sites is the width of the interface. Hence one has $d=30$ ($R_1=10$, $R_2=10$) for $h=1$, $d=40$ ($R_1=20$, $R_2=10$) for $h=2$ and $d=60$ ($R_1=40$, $R_2=10$) for $h=4$. At increasing values of $h$, horizontal lines are crossed more frequenty, an indication that cores come into close contant for longer periods of time. (c) In this plot $T_{int}$ is the interaction time, namely the time (in simulation units) in which a thin film is sustained by the two cores, an effectt occurring when $|\Delta {\bf r}_{cm}|<d$. Also, $T_{tot}=10^6$ timesteps is the total simulation time of a single run. At increasing values of $h$, $T_{int}/T_{tot}$ augments in all regimes of $\dot{\gamma}$ explored.}
\label{fig7} 
\end{figure*}

In spite of the approximate evaluation, the computation of $|\Delta{\bf r}_{cm}|$ unveils a neat result. Indeed, in Fig.\ref{fig7}a-b the time evolution of $|\Delta{\bf r}_{cm}|$ at low shear (a) and high shear (b) for three values of polydispersity index shows that increasing $h$ favours close contacts among cores, since in these regimes $|\Delta {\bf r}_{cm}|$ falls below the distance $d$  for longer periods of time. In Fig.\ref{fig7}a, for example, if $h=1$ the cores come close enough only at late times, whereas if $h=2$ (and similarly for $h=4$) they periodically approach  ($|\Delta {\bf r}_{cm}|\le d$) and detach ($|\Delta {\bf r}_{cm}|>d$) from early times on. In Fig.\ref{fig7}b, at low values of $h$, the cores follow a a dynamic behavior described in Fig.\ref{fig5}e-g, in which they stabilize far apart and collisions are neglibible. If $h=4$, on the contrary, the behavior is essentially that reported in Fig.\ref{fig5}a-c, in which the small core repeatedly moves around the large one at very close distance. Finally in Fig.\ref{fig7}c we plot the time $T_{int}$ spent by the cores in close contact divided by the total simulation time $T_{tot}$ (equal to $10^6$ timesteps) for different values of $h$. In all shear regimes $T_{int}$ grows either gently, at low/moderate values of $\dot{\gamma}$, or abruptly at high $\dot{\gamma}$. The temporary plateau observed at low shear rates (red/plusses plot) for $2<h<3$ is due to the progressive shift of the large drop located in the middle of the emulsion towards the external interface, an effect, described in Fig.\ref{fig2}, caused by the drag exherted by the shear flow. At high shear rates (blue/asterisks plot), the flat curve for $h<2.5$ captures the dynamics described in Fig.\ref{fig4}e-g in which the contacts among cores are essentially absent, while the plateau for $h>3$ accounts for a regime in which, at high $h$, the time spent in close contact  remains basically constant. 

\section{Conclusions}

In conclusion we have numerically investigated the dynamics of a 2D polydisperse double emulsion under a symmetric shear flow in a geometry inspired to those used in realistic microfluidic experiments. As in previous studies \cite{tiribocchi_pof,tiribocchi_prf} we have explored regimes in which the emulsion  undergoes relevant shape deformations, although not sufficient to yield breakup. We have considered the simplest example of polydisperse double emulsion, made of a large drop containing two further drops of different size each. 

Our results suggest that, besides the shear rate $\dot{\gamma}$, the polydispersity index $h$ as well considerably affects the response of the emulsion under shear. At low values of $\dot{\gamma}$, a monodisperse (i.e. $h=1$) double emulsion may attain a state in which one core periodically rotates around a second one remaining essentially motionless in the middle of the emulsion. By increasing the polydispersity ratio, this Earth-Sun-like scenario is replaced by one in which both cores exhibit a periodic motion along elliptical trajectories within the external drop, basically in agreements with previous works \cite{tiribocchi_pof,tiribocchi_prf}. At high values of $\dot{\gamma}$, on the contrary, the large core only weakly fluctuates in the middle of the emulsion while the small one rapidly turn around in a periodic manner, thus re-establishing, to some extent, the Earth-Sun-like dynamics. Decreasing $h$ stabilizes the system in a state in which the two inner cores arrange far apart at opposite sides of the emulsion. 

Polydispersity has been also found to crucially condition interactions among cores and shape deformations of the extenal droplet. Indeed, high values of $h$ may easily favour close contact and collisions among cores when under flow, a situation less likely if polydispersity decreases. This is evaluated in terms of capability of the cores to sustain temporarily fluid films within simulation times long enough to allow the emulsion to achieve a near steady state.  Also, in a regime of high shear rates, morphological deformations of the external shell are found to be mitigated by the presence of a larger core, whose shape is generally less affected by the fluid flow.
We finally note that these results hold as long as the area fraction occupied by the cores remains approximately between $0.01$ and $0.5$. The study of the physics over such limits is part of future investigation, in particular in foamy-like mixtures, where the very high density of the dispersed phase can favour the rupture of the fluid films and the merging of the droplets (hence increase of polydispersity ratio), thus decisively modifying, often {\it en route}, the mechanical properties of the material. 

Alongside polydispersity, the dynamics of double emulsions is also affected by viscosity, surface tension and concentration of the surfactant solution. While increasing the viscosity of the middle fluid, for example, would harden the emulsion \cite{utada2005monodisperse}, augmenting the surface tension would prevent large shape deformations \cite{shardt2013}. A very high value of surface tension could diminish the capillary number towards a threshold below which coalescence can occur, since a flat film could not form to delay droplet merging. Such effect is generally minimized by including a suitable surfactant adsorbed onto the interfaces of the droplets although, in the presence of heavy multi-body collisions, merging becomes more likely. Following previous works \cite{tiribocchi_pof,tiribocchi_nat}, in our simulations we have considered, for simplicity, fluid components (dispersed and continuous phase) with equal viscosity, an approximation that could be released by assuming a suitable functional form of $\eta=\eta(\phi)$. In addition, in our model coalescence is essentially forbidden since repulsive interactions among cores are present. It would be interesting to investigate the physics when only a portion of the interfaces is covered by a surfactant, a situation resembling the case of Janus particles alternatively exposed to surfactant-rich and surfactant-poor fronts.

\section*{Acknowledgments}
A. T., A. M., F. B., M. L. and S. S. acknowledge funding from the European Research Council under the European Union's Horizon 2020 Framework Programme (No. FP/2014-2020) ERC Grant Agreement No.739964 (COPMAT).

\section*{Data Availability Statement}
Data available on request from the authors.

\bibliography{bibliography}

\end{document}